\documentclass[useAMS,usenatbib]{mn2e}

\usepackage{graphicx}

\title[Upper limits on bolometric luminosities of ten type Ia supernova progenitors from \textit{Chandra} observations]{Upper limits on bolometric luminosities of ten type Ia supernova progenitors from \textit{Chandra} observations}
\author[M.T.B. Nielsen, R. Voss and G. Nelemans]{M.T.B. Nielsen$^{1}$\thanks{E-mail:
m.nielsen@astro.ru.nl}, R. Voss$^{1}$ and G. Nelemans$^{1}$\\
$^{1}$Department of Astrophysics, IMAPP, Radboud University Nijmegen, PO Box 9010, NL-6500 GL Nijmegen, the Netherlands}
\begin{document}

\date{Accepted -. Received \today; in original form -}

\pagerange{\pageref{firstpage}--\pageref{lastpage}} \pubyear{2011}

\maketitle

\label{firstpage}

\begin{abstract}
We present an analysis of \textit{Chandra} observations of the position of ten nearby ($<$ 25 Mpc) type Ia supernovae, taken before the explosions. No sources corresponding to progenitors were found in any of the observations. We calculated upper limits on the bolometric luminosities of the progenitors assuming black-body X-ray spectra with temperatures of 30-150 eV. This is inspired by the fact that luminous super-soft X-ray sources have been suggested as the direct progenitors of type Ia supernovae. The upper limits of two supernovae in our sample are comparable to the luminosities of the brightest observed super-soft sources, ruling out such sources as the progenitors of these supernovae. In contrast to \citet{Liu.et.al.2012} we find that for SN2011fe we can rule out Eddington luminosity systems for black body temperatures as low as 40 eV. Our findings are consistent with statistical studies comparing the observed type Ia supernova rate to the number of super-soft sources or the integrated X-ray luminosity in external galaxies. This suggest that either the progenitors of type Ia supernovae are not accreting, nuclear burning white dwarfs, or that they do not look like the classical super-soft sources, e.g. because they are obscured.
\end{abstract}

\begin{keywords}
binaries: close -- supernovae: general -- white dwarfs -- X-rays: binaries
\end{keywords}

\section{Introduction} \label{Sect:Introduction}
Type Ia supernovae (SNe) are important astrophysical phenomena, both in relation to cosmology \citep{Riess.et.al.1998,Perlmutter.et.al.1999} and galactic chemical and dynamical evolution. Despite some fourty years of research on the subject the exact nature of the progenitor systems of SNe Ia remain undetermined. Two scenarios are usually considered by the community (e.g. \citealt{Hillebrandt.Niemeyer.2000}): the single degenerate (SD) and double degenerate (DD) scenarios. In the former, a single carbon-oxygen white dwarf (WD) star accretes matter from a non-degenerate companion star \citep{Whelan.Iben.1973,Nomoto.1982}, thereby growing in mass until it reaches a critical mass ($\sim$ 1.37 M$_{\odot}$), at which point the temperature and density in its interior are high enough for carbon and oxygen to fuse explosively into radioactive iron-group elements. In the DD scenario, two WDs with individual masses less than the Chandrasekhar mass merge to form a single carbon-oxygen WD at or above the critical mass needed for thermonuclear runaway \citep{Tutukov.Yungelson.1981,Webbink.1984,Iben.Tutukov.1984}. In both cases, the resulting SN explosion completely unbinds the WD, and the subsequent decay of radioactive nickel powers a light curve that can be used as standardizable cosmology candles \citep{Philips.1993}.

It has been suggested that the steady accretion and nuclear burning of material on the surface of the WD in the SD scenario will emit super-soft X-rays \citep{van.den.Heuvel.et.al.1992,Kahabka.van.den.Heuvel.1997}. The spectrum of this type of emission is expected to resemble a black-body with $kT_{BB}$= 30-100 eV and luminosities between $10^{37}-10^{38}$ erg s$^{-1}$. For SNe closer than $\sim$ 25 Mpc such emissions should theoretically be observable with the \textit{Chandra} X-ray Observatory. For this reason, a search of archival \textit{Chandra} images taken before the SN explosions was conducted by \citet{Voss.Nelemans.2008}, and the result was one possible detection (SN2007on, however, see also \citealt{Roelofs.et.al.2008}) and four upper limits (SN2002cv, SN2004W, SN2006mr and SN2007sr, see \citealt{Nelemans.et.al.2008}). Upper limits for the progenitor of SN2011fe based on archival \textit{Chandra} images were reported by \citet{Butler.et.al.ATEL3587.2011} and later \citet{Li.et.al.2011}, both studies using a black-body temperature of $kT_{BB}=$ 67 eV. Upper limits for SN2011fe were also reported by \citet{Liu.et.al.2012}, however, below we show that we do not find the same upper limits to the bolometric luminosities as reported in that study.

The search for progenitors in archival X-ray images was inspired by the analoguous search for the progenitors of core-collapse SNe in \textit{HST} archive, see review by \citet{Smartt.2009}. A similar search for SN type Ia progenitors in \textit{HST} archival images was performed by \citet{Mannucci.Maoz.2008} for SN2006dd and SN2006mr in NGC 1316, but no optical counterparts of these SNe were observed. Additionally, \textit{HST} archival images were used to put upper limits on the optical luminosity of the progenitor of SN2007on \citep{Voss.Nelemans.2008} and SN2007sr \citep{Nelemans.et.al.2008}. Limits on the optical magnitude and bolometric luminosity of the progenitor of SN2011fe were reported by \citet{Li.et.al.2011}.

In this paper we present a homogenous analysis of ten recent, nearby ($\leq 25$ Mpc) type Ia SNe for which pre-explosion Chandra images are available: SN2002cv (Larionov et al. (IAUC 7901, IAUC 7903), classified by Meikle \& Matilla (IAUC 7911)), SN2003cg (Itagaki et al.; Arbour (IAUC 8097), classified by Kotak et al. (IAUC 8099)), SN2004W (Moore \& Li, classified by Fillipenko et al. (IAUC 8286)), SN2006X (Suzuki; Migliardi (IAUC 8667), classified by Quimby et al. (CBET 393)), SN2006dd (Monard (CBET 533), classified by Salvo et al. (CBET 557)), SN2006mr (Monard (CBET 723), classified by Phillips et al. (CBET 729)), SN2007gi (Itagaki (CBET 1017), classified by Harutyunya et al. (CBET 1021)), SN2007sr (Drake et al. (CBET 1172), classified by Naito et al. (CBET 1173)), SN2008fp (Pignata et al. (CBET 1506), classified by Wang et al. (CBET 1509)) and SN2011fe (discovered and classified by Nugent et al. (CBET 2792)). We derive upper limits on the bolometric luminosities of the progenitors assuming black-body spectra with effective temperatures between 30 and 150 eV. The luminosities found in this study are compared to those of known super-soft X-ray sources (SSS) in nearby galaxies.

Together with SN2007on the ten SNe examined in this study comprise the complete set of currently-known type Ia SNe that have pre-explosion images in the \textit{Chandra} archive. We note that it is currently unclear if the progenitor of SN2007on has been directly observed or not, see \citet{Voss.Nelemans.2008} and \citet{Roelofs.et.al.2008}. Due to this ambiguity we refrain from dealing with SN2007on in this study.

In Section \ref{Sect:Observations} we describe the \textit{Chandra} observations used in this study. Section \ref{Sect:Data.Reduction} relates the methods employed in the data analysis of these observations. Section \ref{Sect:Discussion} discusses our results, and Section \ref{Sect:Conclusion} concludes.

\section{Observations} \label{Sect:Observations}
By searching the \textit{Chandra} Data Archive we found pre-explosion ACIS-S images at the positions of ten nearby ($<$25 Mpc) type Ia SNe. The SNe in question are SN2002cv, SN2003cg, SN2004W, SN2006X, SN2006dd, SN2006mr, SN2007gi, SN2007sr, SN2008fp and SN2011fe. No obvious sources were found on the pre-explosion images for any of these SNe.

For SN2002cv, SN2003cg, SN2004W, SN2006X, SN2006dd, SN2006mr, SN2007gi and SN2008fp, only a single pre-explosion \textit{Chandra} image exists for each of the SNe, and SN2006dd and SN2006mr are on the same image. Several of these images have long ($>$30 ks) exposure times. For SN2007sr and SN2011fe multiple pre-explosion \textit{Chandra} images exist, and these can be combined to give very long (several 100 ks) exposure times.

The observations analysed in this study are summarised on Table \ref{Table:Observations}.

\section{Data Reduction} \label{Sect:Data.Reduction}
We analysed the \textit{Chandra} observations using the CIAO 4.3 software suite. Initially, we examined the images in the entire photon energy range of \textit{Chandra}, i.e. $\sim$300 eV to $\sim$10 keV, to determine whether a source was present. Thereafter, we limited our analysis to photon energies between 300 eV and 1 keV. For a SSS any counts above 1 keV will be background anyway, so this approach allows more stringent upper limits to be placed on an assumed super-soft progenitor. No sources were found at or near the positions of any of the SNe analysed in this study in the two energy ranges used (i.e. 300 eV - 10 keV and 300 eV - 1 keV).

For our data model we assumed an absorbed black-body, using the spectral models {\tt xsphabs} and {\tt xsbbody}, which correspond to XSPEC's {\tt phabs} and {\tt bbody}, respectively. We generated spectral-weights files for the appropriate interstellar absorption columns (see below) and four different effective temperatures: $kT_{BB}=$ 30 eV, 50 eV, 100 eV, and 150 eV. The spectral-weights files were used to generate exposure maps for each of the images for each of the four effective temperatures. For SN2007sr and SN2011fe multiple pre-explosion images exist, and for these SN position we combined the binned images and the exposure maps to obtain deeper observations.

The distances to the progenitors of SN2002cv, SN2003cg, SN2004W, SN2006X, SN2007gi and SN2008fp were assumed identical to the galactocentric distances to their host galaxies\footnote{host galaxies were obtained from IUA Central Bureau for Astronomical Telegrams online list of SNe http://www.cbat.eps.harvard.edu/lists/Supernovae.html} as listed in the NED online database\footnote{http://ned.ipac.caltech.edu/forms/byname.html}. The distance to the progenitor of SN2006dd and SN2006mr is taken from \citet{Stritzinger.et.al.2010}. The distance to the progenitor of SN2007sr is the one given in \citet{Schweizer.et.al.2008}. For SN2011fe we used the recent distance value given in \citet{Shappee.Stanek.2011}.

The hydrogen columns were either found directly in the literature, or by using the formula \citep{Guver.Ozel.2009}: $N_H = 2.21\cdot10^{21} A_V$ where $N_H$ is the neutral hydrogen column in $\mathrm{cm}^{-2}$, and $A_V$ is the total V-band extinction given in magnitudes. The total V-band extinction is found from the reddening law $R_V = A_V E(B-V)$, where $R_V$ is the reddening, and $E(B-V)$ is the selective optical extinction or color excess ($E(B-V) \equiv A_B - A_V$).

For two SNe (SN2004W and SN2008fp) no explicit values for the hydrogen column, reddening, or extinction could be found in the literature. For the two SNe in NGC1316 the column in the host galaxy was assumed to be negligible, following \citet{Stritzinger.et.al.2010}. For these four cases we used the value for the Galactic column found in \citet{Dickey.Lockman.1990}, as referenced with CIAO's COLDEN tool. For SN2011fe \citet{Chomiuk.et.al.2012} estimated a column value about twice that of the Galactic one, while \citet{Stritzinger.et.al.2010} assumed the column in the host galaxy to be negligible. Since SN2011fe is the closest SN Ia in several decades, and the closest to have pre-explosion archival \textit{Chandra} data, we consider both column values in our analysis. The host galaxies, distances, and columns for the SNe analysed in this study are summarised in Table \ref{Table:Host.Gals.Dist.Columns}.

\begin{table*}
 \begin{minipage}{1.\textwidth}
\caption{Host galaxies, distances and total hydrogen columns for each of the SNe analyzed in this study.}
 \centering
  \begin{tabular}{@{}c c c c c c @{}}
  \hline
  supernova & host	& distance	& absorbing		& reference	 		\\
	    & galaxy	& [Mpc]		& column		&				\\
	      &		&		& [$n_H$ cm$^{-2}$]	&				\\
  \hline
  \hline
  2002cv & NGC 3190 	& 16.4		& $1.93\cdot10^{22}$	& \citet{Elias-Rosa.et.al.2008}	\\
  \hline
  2003cg & NGC 3169 	& 15.1		& $2.99\cdot10^{21}$	& \citet{Elias-Rosa.et.al.2006}	\\
  \hline
  2004W  & NGC 4649	& 14.6		& $2.12\cdot10^{21}$	& \citet{Dickey.Lockman.1990}	\\
  \hline
  2006X  & NGC 4321	& 20.9		& $2.30\cdot10^{21}$	& \citet{Wang.et.al.2008}	\\
  \hline
  2006dd & NGC 1316	& 17.8 		& $2.13\cdot10^{20}$	& \citet{Dickey.Lockman.1990}	\\
  \hline
  2006mr & NGC 1316	& 17.8 		& $2.13\cdot10^{20}$	& \citet{Dickey.Lockman.1990}	\\
  \hline
  2007gi & NGC 4036	& 21.2 		& $6.85\cdot10^{20}$	& \citet{Zhang.et.al.2010}	\\ 
  \hline
  2007sr & NGC 4038/39	& 22.3 		& $4.00\cdot10^{20}$	& \citet{Nelemans.et.al.2008}	\\
  \hline
  2008fp & ESO 428-G14	& 20.4		& $2.21\cdot10^{21}$	& \citet{Dickey.Lockman.1990}	\\
  \hline
  2011fe & M 101	& 6.4		& $3.02\cdot10^{20}$/$1.14\cdot10^{20}$	& \citet{Chomiuk.et.al.2012}/\citet{Dickey.Lockman.1990}\\
  \hline
\end{tabular} \label{Table:Host.Gals.Dist.Columns}
\end{minipage}
\end{table*}

For each observation we found a good estimate of the average number of background photons from a suitably-chosen region free of point-sources close to the source. We then used a circular aperture of radius 4.5 pixels (covering more than 90\% of the point-spread function of all observations) around the position of the SN and extract the number of photons. This aperture contains a Poissonian realisation of the expected average number of counts from a source plus the background\footnote{This was unproblematic for all observations except one; for SN2006mr choosing a suitable background region was more difficult, due to the proximity of a large, unresolved and uneven background. To be conservative, for SN2006mr we chose a background region that was less bright than the immediate background of the assumed progenitor position and made our calculations using the resulting background photon count. Since this background is clearly smaller than the actual background our upper limits for the progenitor of SN2006mr should be considered even more solid than the rest of our results. However, it also means that formally our analysis indicates the presence of a source at the position of the progenitor of SN2006mr, even though it is not actually possible to infer the presence of one.}. For this photon count $N_{\mathrm{obs}}$ we found the maximum average number of counts $\mu$, for which the probability $P$ of observing $N_{\mathrm{obs}}$ photons is within 3$\sigma$, assuming Poissonian statistics, see e.g. \citet{Gehrels.1986}: $P \left( \mu, N \leq N_{\mathrm{obs}} \right) \leq 0.0013$. This $\mu$ represents the 3$\sigma$ upper limit of any progenitor including background. We find the upper limit to the luminosity of the source according to the formula,

\begin{eqnarray}
 L_{X,UL} = 4 \pi \frac{\left( \mu - b \right) \langle E_{\gamma} \rangle d^2}{\zeta}
\end{eqnarray}
where $b$ is the expected background for a circular aperture of radius 4.5 pixels, $\langle E_{\gamma} \rangle$ is the average energy of the photons found from the absorbed XSPEC model for the assumed spectrum, $d$ is the distance to the SN and $\zeta$ is the value of the exposure map for the given spectrum at the position of the SN on the detector.

The luminosities were then corrected for interstellar absorption to yield the unabsorbed luminosities.

Since our data model is limited to photons from 300 eV to 1 keV the luminosities found are scaled to provide bolometric luminosities,
\begin{eqnarray}
L_{bol,UL} =  \frac{L_{X,UL}}{C} \label{Eq:Bolometric.Correction}
\end{eqnarray}

For the values of $kT_{BB}$ used in our analysis the scaling factors are:
\begin{eqnarray}
30eV  &:& C = 9.58\cdot10^{-3}  \nonumber \\
50eV  &:& C = 1.40\cdot10^{-1}  \nonumber \\
100eV &:& C = 6.01\cdot10^{-1}  \nonumber \\
150eV &:& C = 7.22\cdot10^{-1}  \nonumber \\
\end{eqnarray}

\begin{figure}
 \hspace{0pt}\mbox{
 \begin{minipage}[c]{.48\textwidth}
  \centering
  \includegraphics[width=0.75\textwidth]{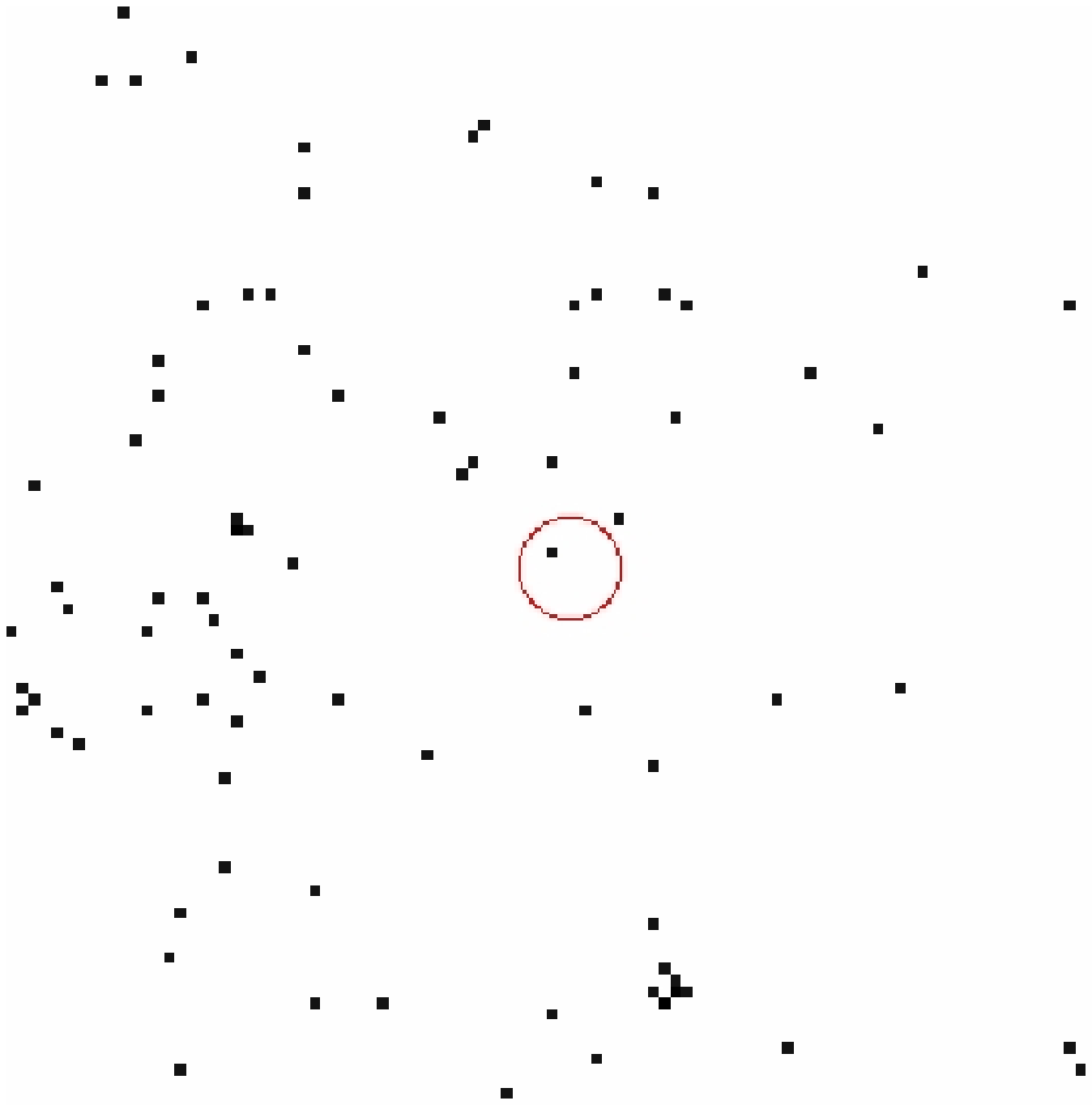}
  \caption{Cut-out of \textit{Chandra} image for observation 2760. The circle corresponds to an aperture of 4.5 pixels at the position of SN2002cv.}\label{Fig:SN2002cv}
 \end{minipage}
 \hspace{.03\textwidth}
 \begin{minipage}[c]{.48\textwidth}
  \centering
  \includegraphics[width=0.75\textwidth]{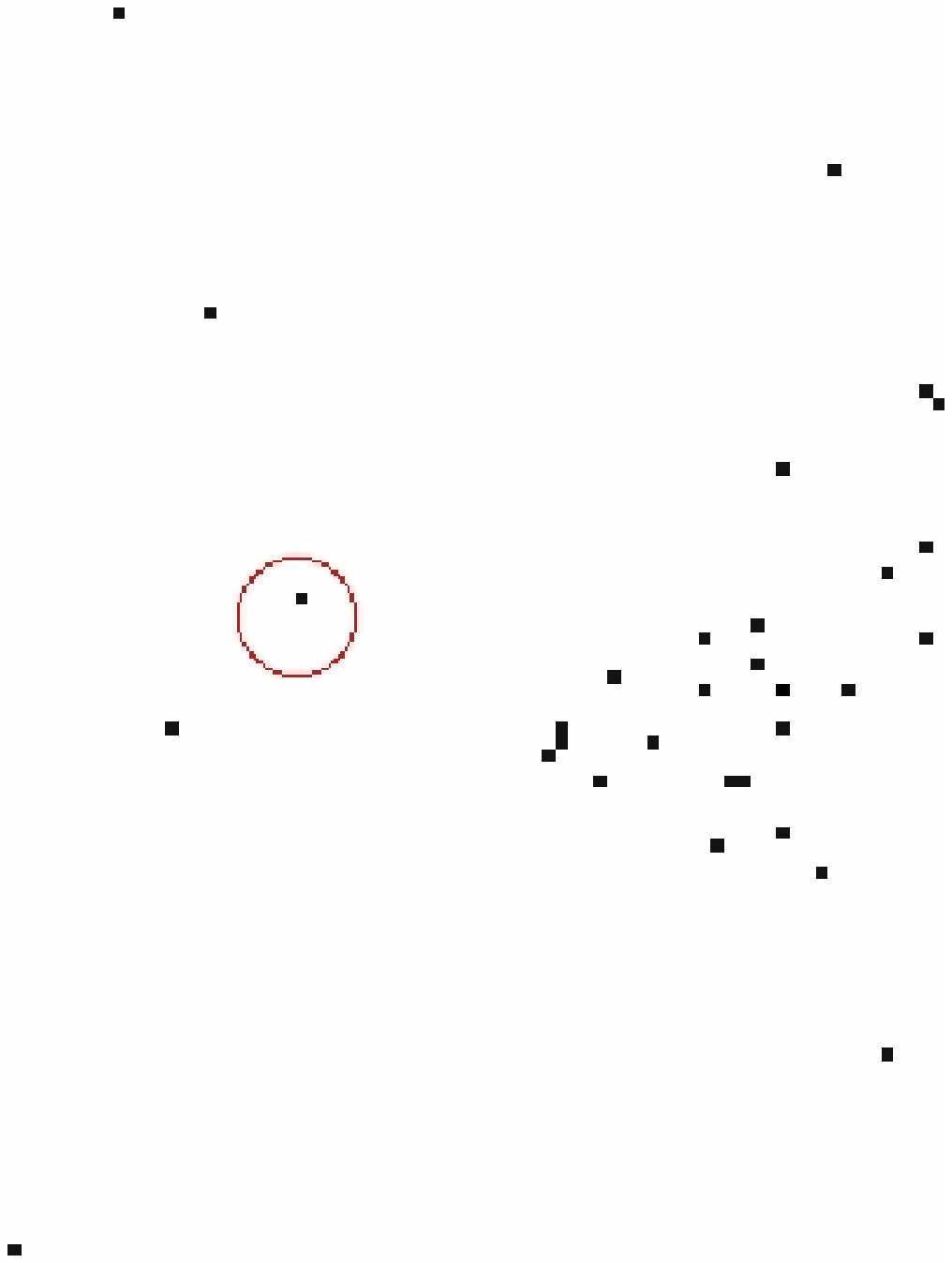}
  \caption{Cut-out of \textit{Chandra} image for observation 1614. The circle corresponds to an aperture of 4.5 pixels at the position of SN2003cg.}
 \end{minipage}
}
 \hspace{0pt}\mbox{
 \begin{minipage}[c]{.48\textwidth}
  \centering
  \includegraphics[width=0.75\linewidth]{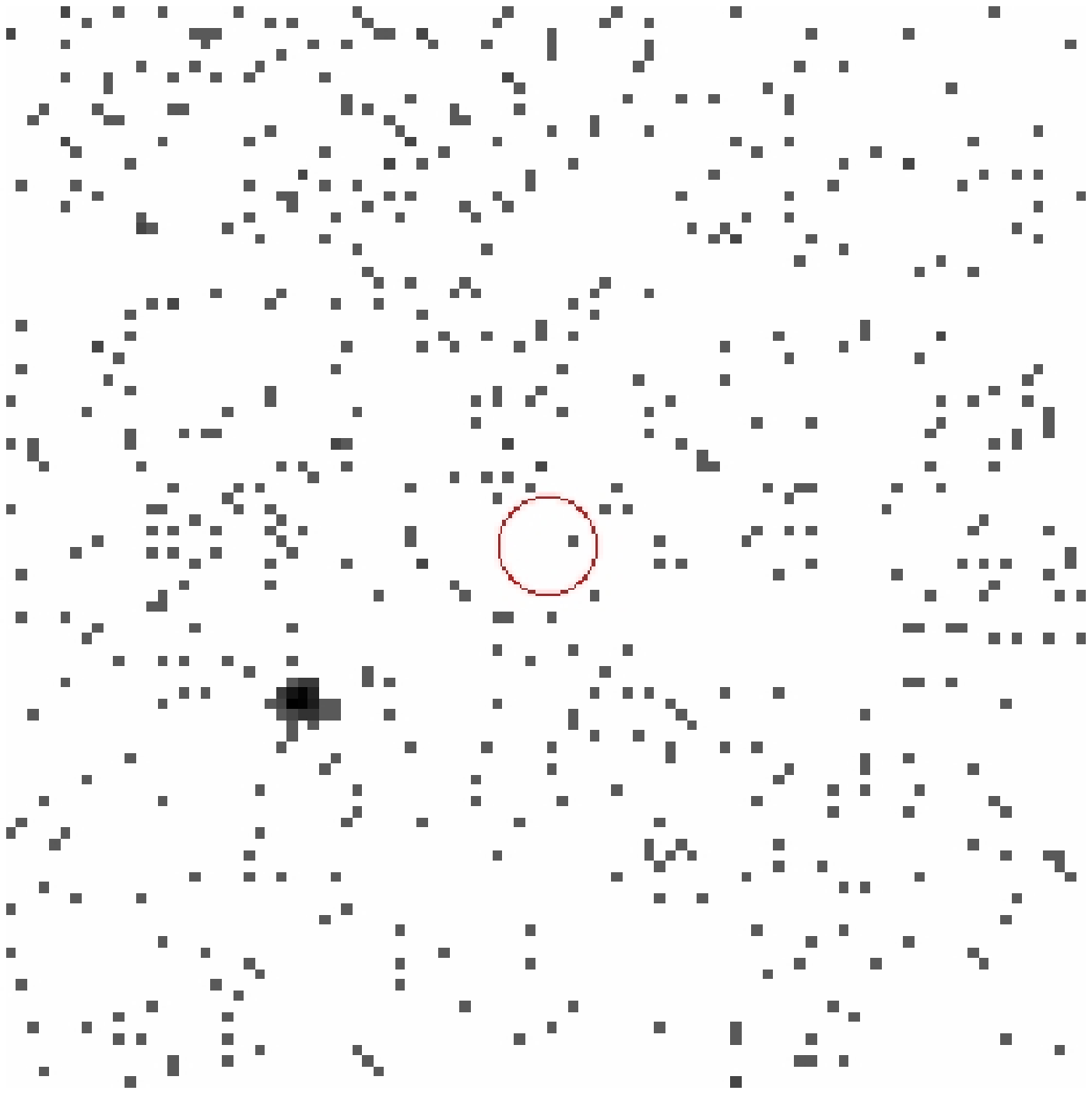}
  \caption{Cut-out of \textit{Chandra} image for observation 785. The circle corresponds to an aperture of 4.5 pixels at the position of SN2004W.}
 \end{minipage}
 \hspace{.03\textwidth}
 \begin{minipage}[c]{.48\textwidth}
  \centering
  \includegraphics[width=0.75\textwidth]{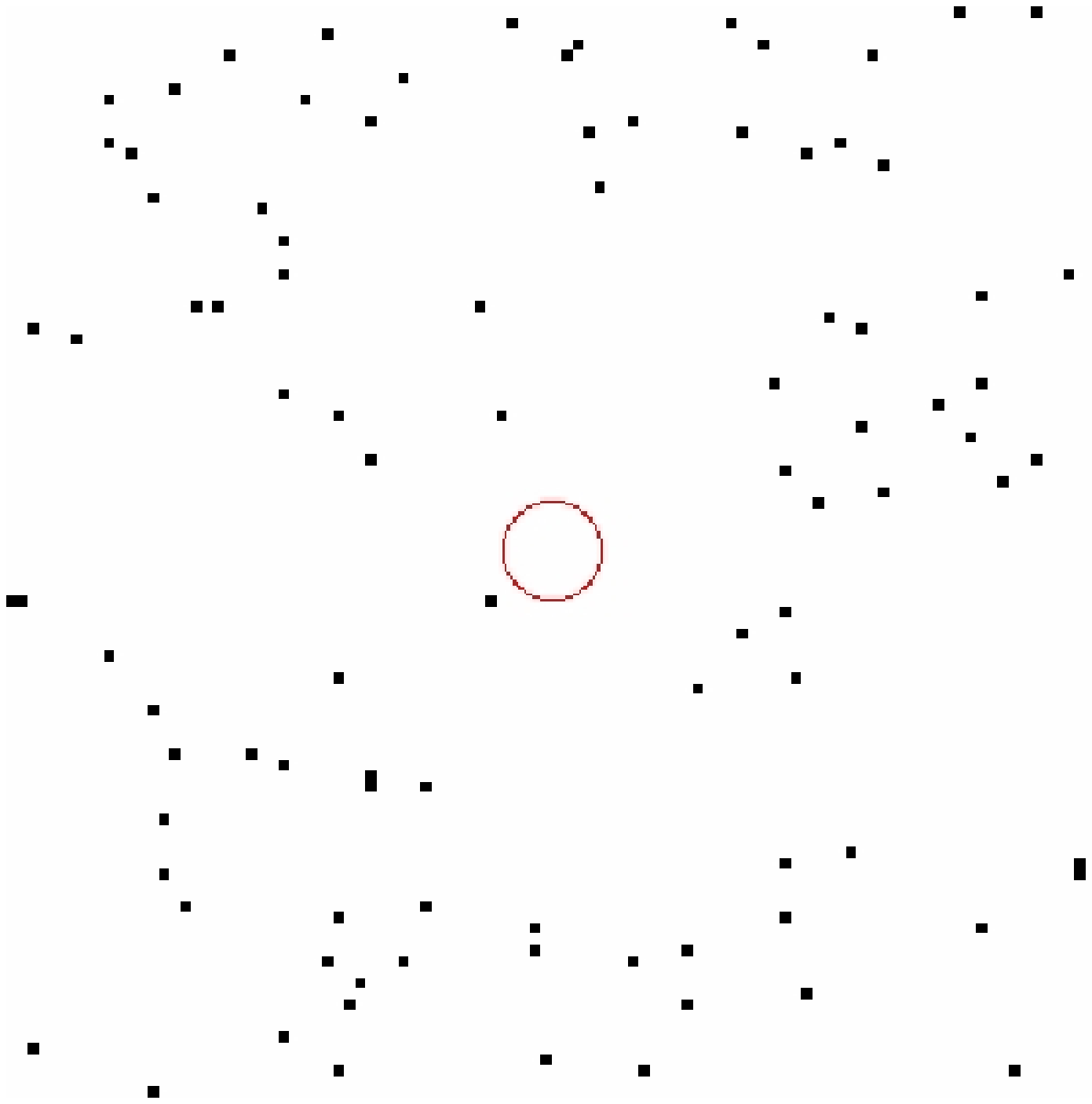}
  \caption{Cut-out of \textit{Chandra} image for observation 400. The circle corresponds to an aperture of 4.5 pixels at the position of SN2006X.}
 \end{minipage}
}
 \hspace{25pt}\mbox{
 \begin{minipage}[c]{.48\textwidth}
  \centering
  \includegraphics[width=0.75\textwidth]{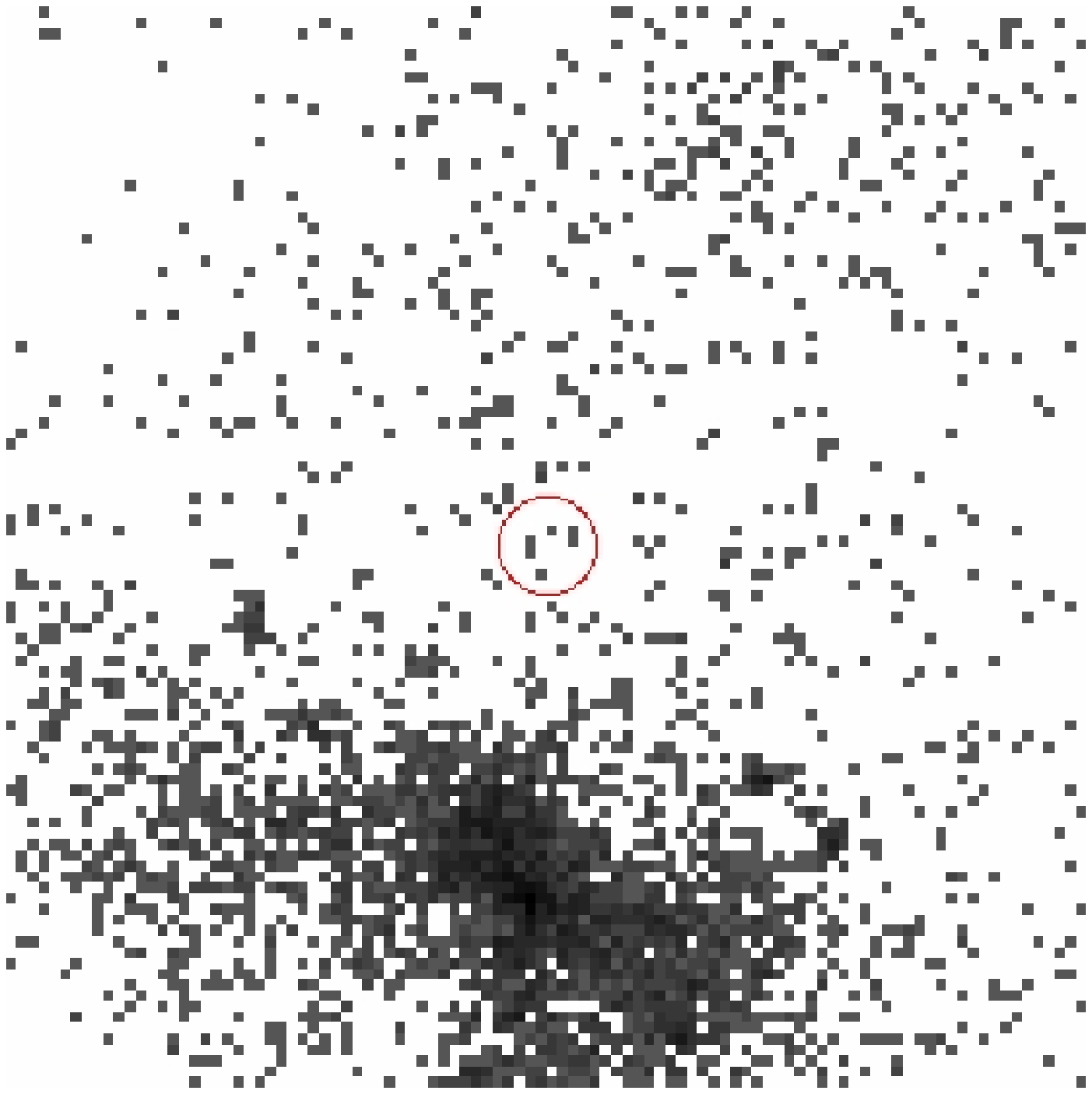}
  \caption{Cut-out of \textit{Chandra} image for observation 2022. The circle corresponds to an aperture of 4.5 pixels at the position of SN2006dd.}
 \end{minipage}
 \hspace{.03\textwidth}
 \begin{minipage}[c]{.48\textwidth}
  \centering
  \includegraphics[width=0.75\textwidth]{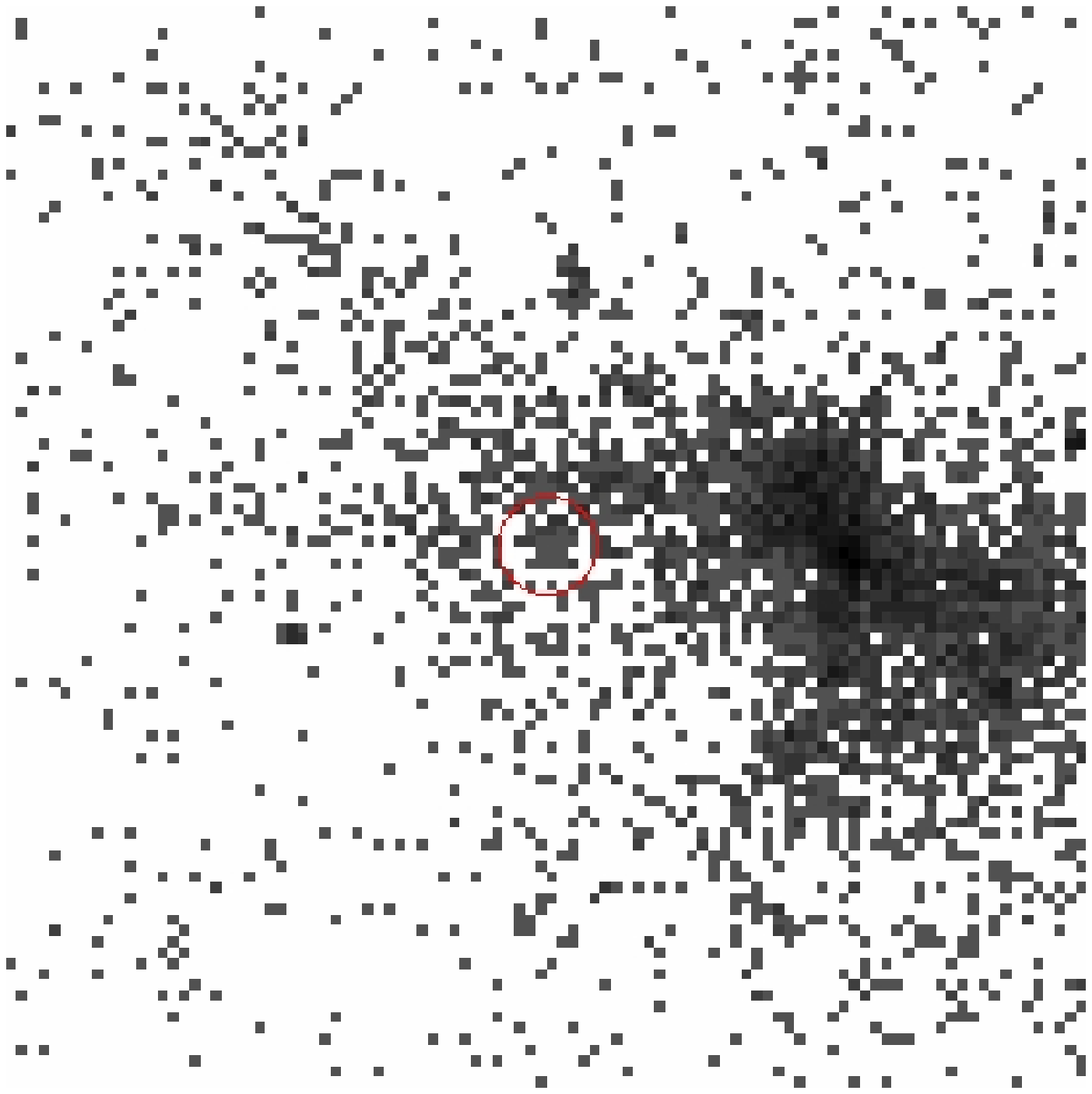}
  \caption{Cut-out of \textit{Chandra} image for observation 2022. The circle corresponds to an aperture of 4.5 pixels at the position of SN2006mr.}
 \end{minipage}
}
\begin{center}
\end{center}
\addtolength{\textwidth}{-85pt}
\end{figure}

\begin{figure}
 \hspace{0pt}\mbox{
 \begin{minipage}[c]{250pt}
  \centering
  \includegraphics[width=0.7\textwidth]{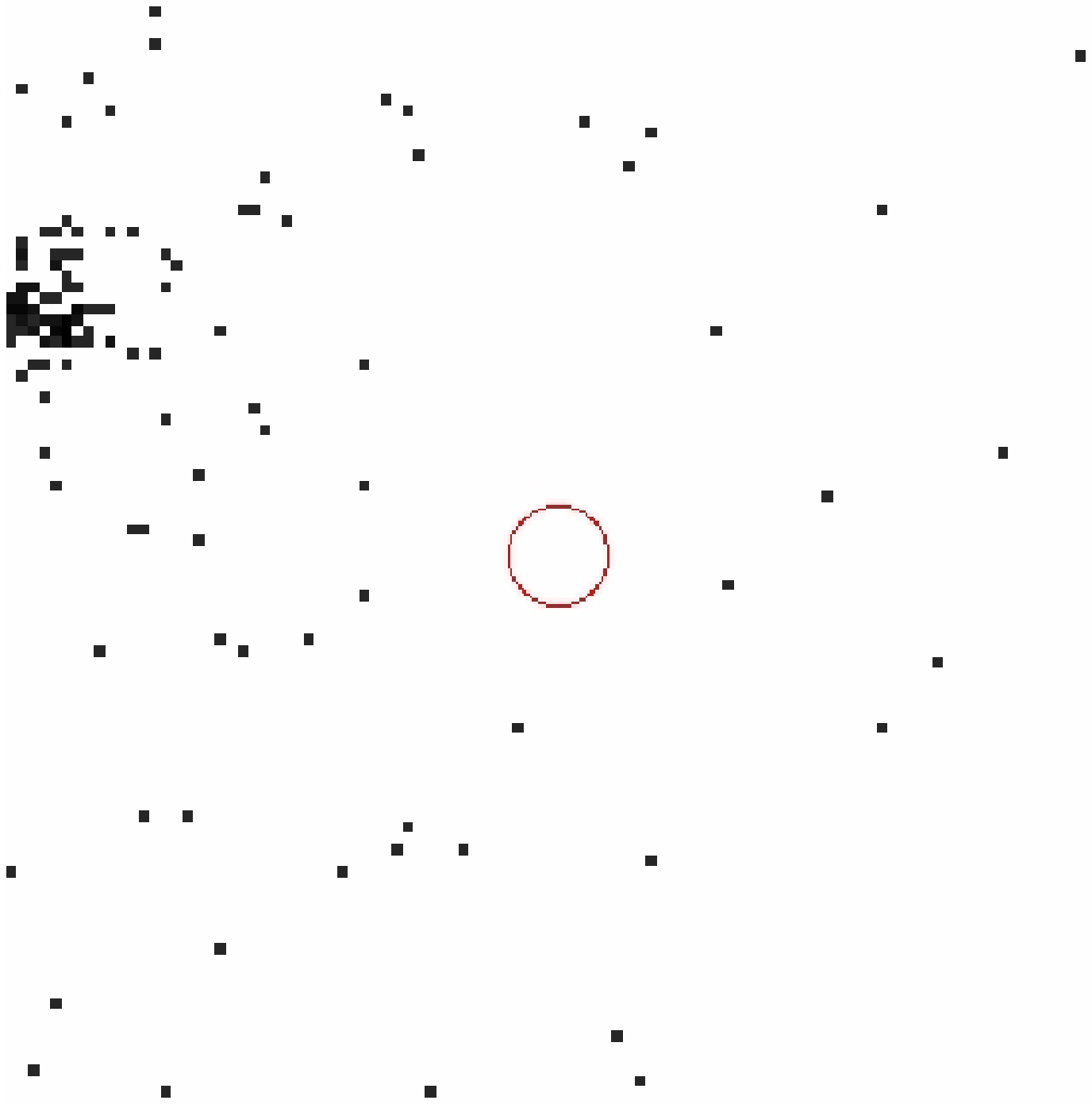}
  \caption{Cut-out of \textit{Chandra} image for observation 6783. The circle corresponds to an aperture of 4.5 pixels at the position of SN2007gi.}
 \end{minipage}
 \hspace{.03\textwidth}
 \begin{minipage}[c]{250pt}
  \centering
  \includegraphics[width=0.75\textwidth]{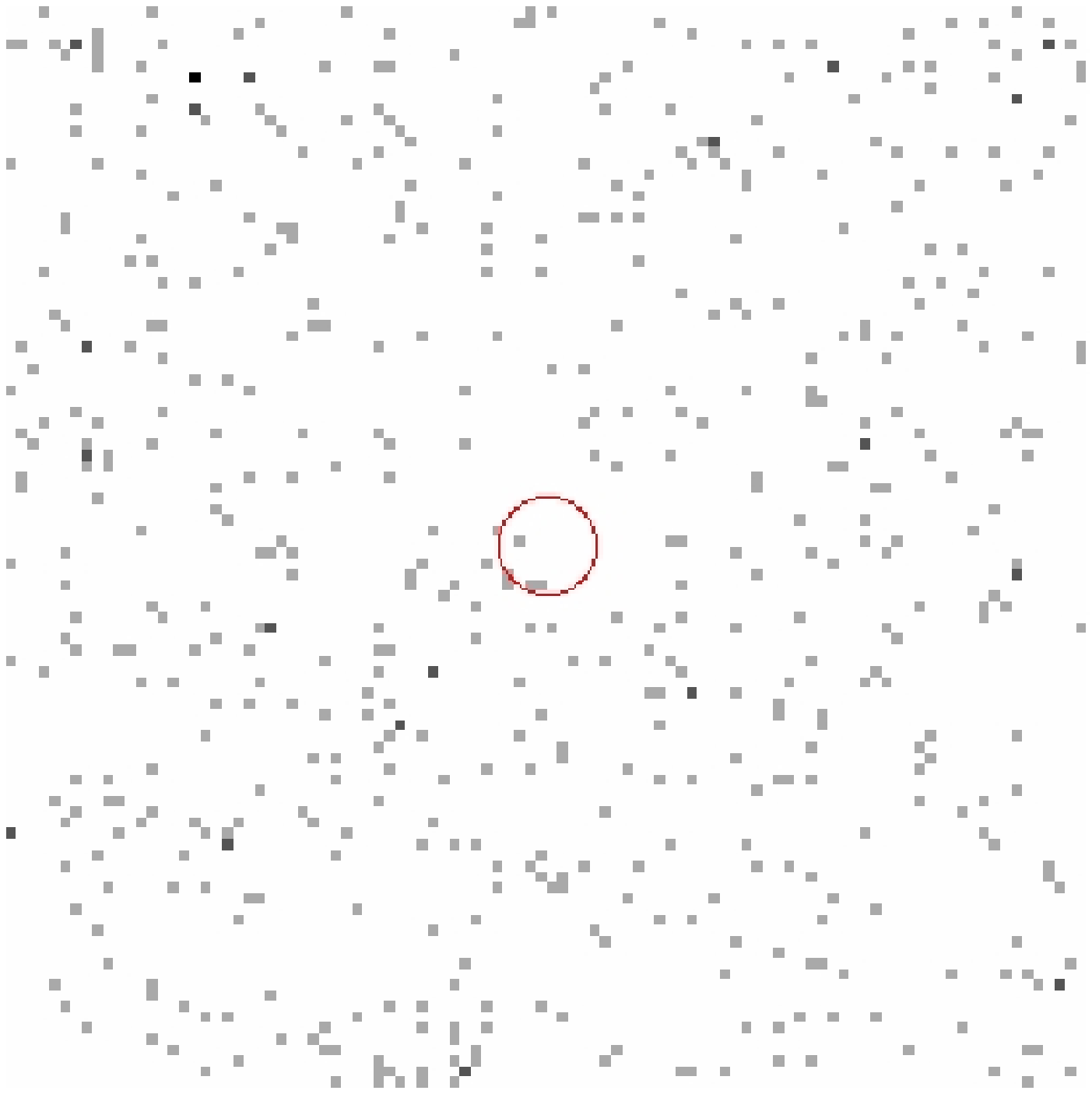}
  \caption{Cut-out of combined image consisting of \textit{Chandra} observations 315 3040 3041 3042 3043 3044 3718. The circle corresponds to an aperture of 4.5 pixels at the position of SN2007sr}
 \end{minipage}
}
 \hspace{0pt}\mbox{
 \begin{minipage}[c]{250pt}
  \centering
  \includegraphics[width=0.75\textwidth]{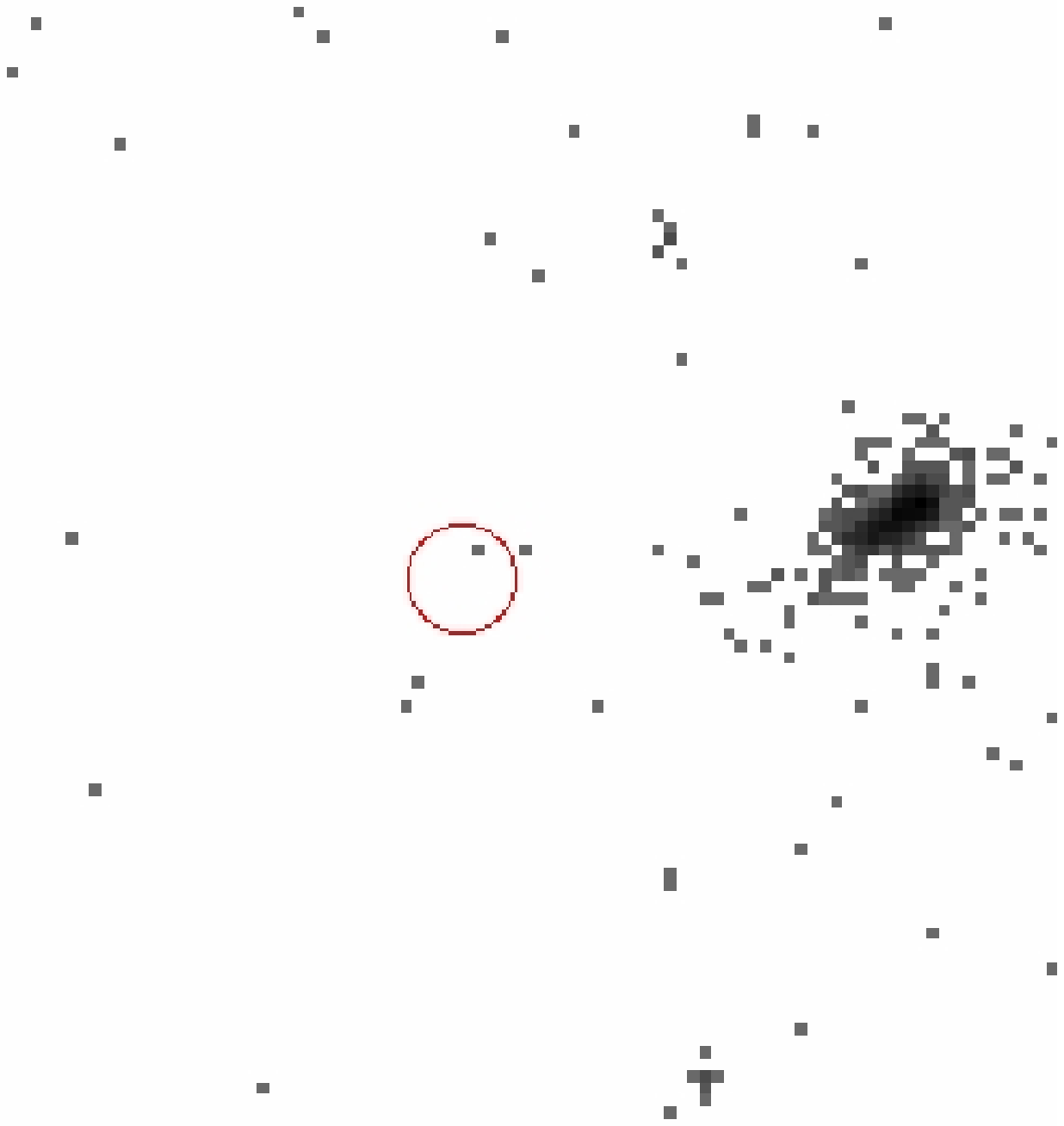}
  \caption{Cut-out of \textit{Chandra} image for observation 4866. The circle corresponds to an aperture of 4.5 pixels at the position of SN2008fp.}
 \end{minipage}
 \hspace{.03\textwidth}
 \begin{minipage}[c]{250pt}
  \centering
  \includegraphics[width=0.7\textwidth]{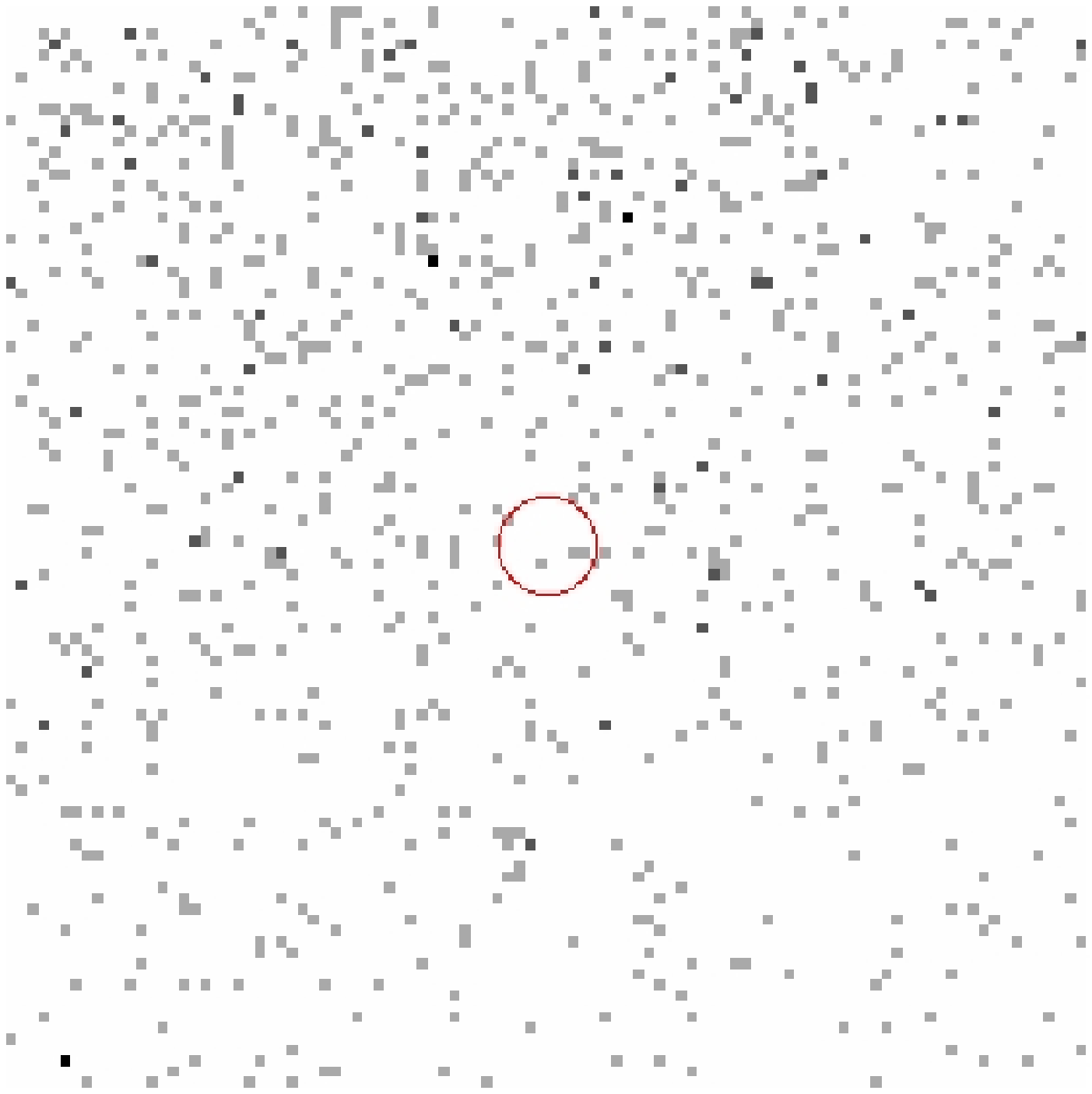}
  \caption{Cut-out of combined image consisting of \textit{Chandra} observations 4731, 5296, 5297, 5300, 4732, 5309, 4733, 5322, 5323, 4737, 5338, 5339, 5340, 4735, 6114, 6115, 6118, 4736, 6152, 4737, 6169, 6170, 6175. The circle corresponds to an aperture of 4.5 pixels at the position of SN2011fe.}\label{Fig:SN2011fe}
 \end{minipage}
}
\begin{center}
\end{center}
\end{figure}

The observations analysed in this study, along with the photon counts and exposure map values used to calculate the upper limit luminosities, are listed in Tables \ref{Table:Observations} and \ref{Table:Upper.Limits}. The individual \textit{Chandra} images are shown on Figures \ref{Fig:SN2002cv}-\ref{Fig:SN2011fe}. These images show all events from 0.3 to 1 keV.

Our results are summarised in Table \ref{Table:Upper.Limits}.

\section{Discussion} \label{Sect:Discussion}
Disregarding the ambiguous case of SN2007on, we now have ten pre-explosion \textit{Chandra} X-ray images of the positions of SNe Ia, and none of them show evidence of a progenitor.

\begin{figure}
 \centering
  \includegraphics[width=1.0\linewidth]{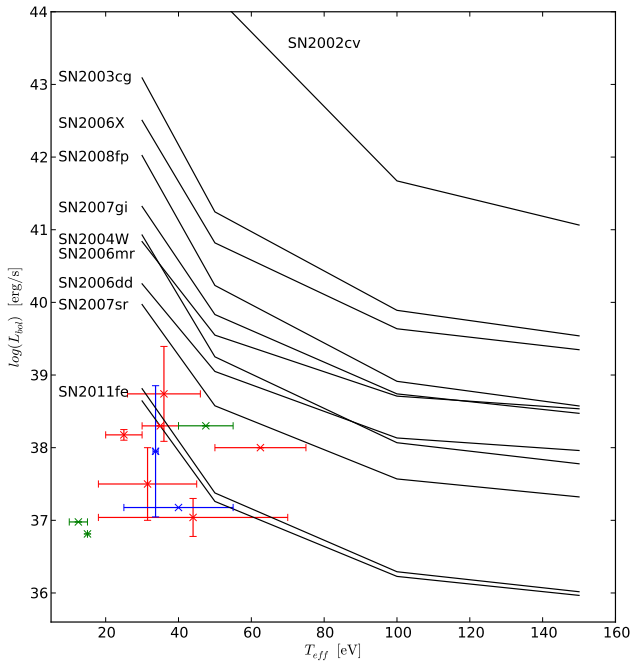} 
 \caption{Comparison between the bolometric luminosity upper limits found in this paper with bolometric luminosities of known super-soft X-ray sources in nearby galaxies. The black lines are the upper limits reported in this paper (for SN2011fe the two curves correspond to the two different column values used, see section \ref{Sect:Data.Reduction}). The green, blue and red points are known persistent SSS in the Milky Way, SMC and LMC, respectively, taken from the online SSS catalog by Greiner, see \citet{Greiner.2000} and references therein. The plot only include sources that are characterized as close-binary super-soft (CBSS) or symbiotic (Sy) systems. We note that many of the fits are based on single observations, and suffer from relatively large systematic uncertainties.}
   \label{Fig:Comparison.Upper.Limits}
\end{figure}

Figure \ref{Fig:Comparison.Upper.Limits} shows a comparison between our results and the bolometric luminosities of known persistent close-binary and symbiotic SSSs in the Galaxy, Large Magellanic Cloud and Small Magellanic Cloud from \citet{Greiner.2000}. Clearly, the bolometric luminosity upper limits of SN2011fe and SN2007sr are probing the luminosity space of 'canonical' SSSs (i.e. $kT_{BB}=$ 30-100 eV, $L_{bol}=10^{37}-10^{38}$ erg/s). For both these SNe we can rule out a naked, bright SSS progenitor. However, a SSS progenitor in the lower part of the expected effective temperature space is still permitted by the observations.

Upper limits based on archival \textit{Chandra} observations were reported previously for SN2002cv, SN2004W and SN2006mr by \citet{Voss.Nelemans.2008} and later corrected in \citet{Nelemans.et.al.2008}. However, the method used in those references was rather different from the one used in the present study: the \textit{Chandra} counts were binned into soft, medium and hard photons, and for each energy bin the total luminosity was calculated from an assumption that the spectrum was flat. The upper limits given for SN2002cv, SN2004W and SN2006mr in \citet{Voss.Nelemans.2008} were $\leq 7.9\cdot10^{37}\mathrm{erg/s}$, $\leq 3.5\cdot10^{37}\mathrm{erg/s}$, and $\leq 1.3\cdot10^{38}\mathrm{erg/s}$, respectively. However, due to the simplified spectral assumption used in that study, and the fact that only Galactic hydrogen was taken into account in \citet{Voss.Nelemans.2008}, the upper limits found in the present study should be considered more accurate. Bolometric upper limits for SN2007sr for $kT_{BB}=$ 50, 100, 150 eV were reported in \citet{Nelemans.et.al.2008} consistent with the results of the present study. For SN2011fe X-ray upper limits of $< 10^{36}$ erg/s were reported by \citet{Butler.et.al.ATEL3587.2011} for photons with energies between 300 and 700 eV and a 67 eV black-body spectral model. Subsequently, \citet{Li.et.al.2011} reported upper limits of $2.7\cdot10^{37}$ erg/s on the bolometric luminosity of the progenitor of SN2011fe, similarly based on a black-body model with $kT_{BB}=$ 67 eV. The results of both of the aforementioned studies are in agreement with the results of this paper. We note that the slightly larger upper limits found by \citet{Li.et.al.2011} can be explained by the shorter exposure time of their combined image used in their study.

We note that upper limits for SN2011fe were also reported in \citet{Liu.et.al.2012}, who found a value of $L_{\mathrm{X}}<6.2\cdot10^{35} \mathrm{erg/s}$ for $kT_{BB}=100 \mathrm{eV}$ in the same energy band as the one used in our study ($0.3-1 \mathrm{keV}$). However, for reasons that are unclear to us \citet{Liu.et.al.2012} subsequently finds different corresponding bolometric luminosities, which could indicate an incorrect conversion factor from X-ray to bolometric luminosity. Additionally, that paper makes an incorrect statement concerning the black-body temperatures and response matrices used to find the upper limits found in our study (last paragraph of Section 3 in that article). As should be clear from our Section \ref{Sect:Data.Reduction} we have consistently used the black-body temperature to calculate our X-ray to bolometric luminosity conversion factors. Also, the fact that the ACIS-S detector is unreliable below 300 eV has no impact on our results, because we only use observed photons with energies above this threshold.

As mentioned in the introduction, \citet{DiStefano.2010a} showed that the number of observed SSSs in nearby galaxies is one to two orders of magnitude too small compared with the estimated number of expected SSSs if these were the progenitors of type Ia SNe. A similar result was found by \citet{Gilfanov.Bogdan.2010} who showed that the integrated super-soft X-ray luminosity of elliptical galaxies is roughly two orders of magnitude too low to account for the SN Ia rate. Although our constraints are not as strong as those found in the two above-mentioned studies our results are consistent with them.

The search for archival \textit{Chandra} images was initially undertaken in an attempt to solve the SD vs. DD question, since at least in the na\"{\i}ve picture the SD progenitors were expected to be X-ray bright, while the DD progenitors were not. However, as a number of recent studies show the question of X-ray brightness of Ia progenitors has turned out to be somewhat more complicated than this:

\textbullet \textit{Chandra} and other current X-ray satellites are only sensitive at photon energies considerably above the $kT_{BB}$ of SSS, above $\sim$ 300 eV or so. The corrections applied in Eq. (\ref{Eq:Bolometric.Correction}) illustrate that a small change in $kT_{BB}$ has a drastic effect on the correctional constant $C$ for lower-$kT_{BB}$ sources. The effective temperature of a SD progenitor depends crucially on the extent of the emitting region, and  the radius of an actual SD accretor therefore does not have to diverge much from that of the theoretical model to make the system unobservable to \textit{Chandra}. For such lower $kT_{BB}$-sources, UV observations should be more useful than X-rays. However, UV observations of these sources are problematic for other reasons, such as interstellar extinction.

\textbullet As discussed by \citet{Hachisu.et.al.2010}, a significant fraction of the progenitors of SD SNe Ia may spend the final phase of their accretion towards going SN in the nova regime where their accretion and associated X-ray emission will be periodic instead of continuous. For recent observations of super-soft X-ray emissions from novae see \citet{Henze.et.al.2010}, \citet{Henze.et.al.2011}, \citet{Schaefer.Collazzi.2010} and \citet{Voss.et.al.2008}.

\textbullet Even a steadily-accreting massive WD consistent with a naked, canonical SSS may be obscured by local matter lost from the system, see \citet{Nielsen.et.al.2011}. Several recent studies have found indications of the presence of circumstellar matter, e.g. \citet{Gerardy.et.al.2004}, \citet{Immler.et.al.2006}, \citet{Borkowski.et.al.2006}, \citet{Patat.et.al.2007}, \citet{Chiotellis.et.al.2011}, \citet{Sternberg.et.al.2011}.

\textbullet If the progenitor is a rapidly rotating WD of the type suggested in \citet{DiStefano.et.al.2011} the X-ray emission of the progenitor would have ceased long before the explosion itself.

\textbullet The detailed spectral shape of the SSS is uncertain \citep{Orio.2006}, and the assumption of a black-body spectrum used in this study may therefore be inaccurate. Due to the higher sensitivity of Chandra above 1 keV the upper limits are more constraining for harder spectra. This is at most an order of magnitude different compared to our 150 eV data points, for the unrealistic assumption of a powerlaw with photon index $\Gamma=2$, typical of X-ray binaries (c.f. the limits for power law and black-body in \citet{Li.et.al.2011} for 2011fe).

\textbullet To make things more difficult, a DD progenitor may also emit soft X-rays for a significant period of time, see \citet{Yoon.et.al.2007}. However, the luminosities expected in this scenario are approximately an order of magnitude lower than for the steadily-accreting SD progenitor. In any case, the detailed workings of the DD merger are still not fully understood (cf. \citealt{Pakmor.et.al.2010}, \citealt{Loren-Aguilar.et.al.2009}, \citealt{van.Kerkwijk.et.al.2010}). It is currently unclear if the lighter WD forms a disk around the more massive companion, or if both WDs are disrupted in the course of the merging event, and this would have an important effect on the possible X-ray emissions from such systems.

For the above-mentioned reasons the question of a SD vs. DD progenitor cannot be decided based solely on the X-ray brightness (or lack hereof) of a type Ia SN progenitor. However, a direct detection of X-ray emissions from a progenitor would still be interesting, and would provide much needed observational evidence for the progenitors of type Ia SNe with which to compare theoretical work, something that is sorely lacking at the moment.

\section{Conclusions} \label{Sect:Conclusion}
We have examined archival \textit{Chandra} pre-explosion images corresponding to the position of ten SNe Ia to determine upper limits to the bolometric luminosities of the progenitors.

Disregarding the ambigious case of SN2007on, our study comprises a complete list of nearby SNe that have pre-explosion images in \textit{Chandra}. We compared this sample with known SSSs in the Milky Way and Magellanic Clouds. While most of the luminosities of our sample SNe are too loosely constrained, two SNe (SN2007sr and SN2011fe) probe the luminosity space of known SSSs. The results indicate that the progenitors of these SNe were not bright SSSs shortly before they exploded as SNe Ia. However, our upper limits are not constraining enough to rule out less-bright super-soft X-ray progenitors.

The theoretical picture concerning the super-soft X-ray characteristics of SN Ia progenitors is less than clear. A non-detection does not rule out a SD progenitor, but neither does a positive detection necessarily implicate a SD progenitor or rule out a DD progenitor. Regardless, the archival search method of the \textit{Chandra} archive is highly useful in putting much-needed observational constraints on the progenitors, and is a powerful complement to statistical studies of the characteristics of progenitor populations. The method will become increasingly useful as the sky coverage grows. As SN2011fe shows, if a SN Ia explodes in a nearby galaxy, the chances that several pre-explosion \textit{Chandra} images of the position exist are good, hence affording stringent upper limits to be calculated, or, in the case of an X-ray bright progenitor, a direct detection to be made.

\section{Acknowledgments} \label{Sect:Acknowledgments}

The authors are grateful to the referee Martin Henze for carefully reading our first draft and providing many useful comments and suggestions.

This research made use of data obtained from the \textit{Chandra} Data Archive and the CIAO 4.3 software provided by the \textit{Chandra} X-ray Center. We also acknowledge the IAU Central Bureau of Astronomical Telegrams for providing their list of SNe.

This research is supported by NWO Vidi grant 016.093.305.

Additionally, we acknowledge Gijs Roelofs for help with this project in its early stages.

\begin{table*}
 \begin{minipage}{140mm}
 \caption{\textit{Chandra} observations used in this study. All observations are with the ACIS-S detector.}
 \centering
  \begin{tabular}{@{}c c c c c c @{}}
  \hline
  \textit{Chandra}	& exposure	& pointing  		& SN 		& SN			& observation \\
  observation	& time			& (RA, DEC)		& covered	& explosion		& date \\
		& [ks]			&			& 		& date			& date \\
  \hline
  \hline
    2760	& 20.07		& (10:18:06.50, +21:49:41.70)	& 2002cv	& 2002-05-13		& 2002-03-14 \\
    1614	& 2.15		& (10:14:15.00, +03:27:57.10)	& 2003cg	& 2003-03-21		& 2001-05-02 \\
    785		& 37.35		& (12:43:40.30, +11:32:58.00)	& 2004W		& 2004-01-28 		& 2000-04-20 \\
    400		& 2.53		& (12:22:54.80, +15:49:20.00)	& 2006X		& 2006-02-04 		& 1999-11-06 \\
    2022	& 30.23		& (03:22:41.70, -37:12:29.00)	& 2006dd	& 2006-06-19 		& 2001-04-24 \\
		&		&				&\& 2006mr	& 2006-11-05 		&	\\
    6783	& 15.13		& (12:01:26.90, +61:53:44.00)	& 2007gi	& 2007-07-31 		& 2006-07-24 \\
    315		& 73.17		& (12:01:53.70, -18:52:35.50)	& 2007sr	& 2007-12-18 		& 1999-12-01 \\
    3040	& 69.93		& (12:01:53.70, -18:52:35.50)	& 2007sr	& 2007-12-18 		& 2001-12-29 \\
    3041	& 73.85		& (12:01:53.70, -18:52:35.50)	& 2007sr	& 2007-12-18 		& 2002-11-22 \\
    3042	& 68.14		& (12:01:53.70, -18:52:35.50)	& 2007sr	& 2007-12-18 		& 2002-05-31 \\
    3043	& 67.96		& (12:01:53.70, -18:52:35.50)	& 2007sr	& 2007-12-18 		& 2002-04-18 \\
    3044	& 36.97		& (12:01:53.70, -18:52:35.50)	& 2007sr	& 2007-12-18 		& 2002-07-10 \\
    3718	& 35.16		& (12:01:53.70, -18:52:35.50)	& 2007sr	& 2007-12-18 		& 2002-07-13 \\
    4866	& 30.16		& (07:16:31.20, -29:19:29.00)	& 2008fp	& 2008-09-11 		& 2003-12-26 \\
    4731	& 56.96		& (14:03:12.90, +54:20:55.60)	& 2011fe	& 2011-08-24 		& 2004-01-19 \\
    5296	& 3.23		& (14:03:12.90, +54:20:55.60)	& 2011fe	& 2011-08-24 		& 2004-01-21 \\
    5297	& 21.96		& (14:03:12.90, +54:20:55.60)	& 2011fe	& 2011-08-24 		& 2004-01-24 \\
    5300	& 52.76		& (14:03:12.90, +54:20:55.60)	& 2011fe	& 2011-08-24 		& 2004-03-07 \\
    4732	& 70.69		& (14:03:12.90, +54:20:55.60)	& 2011fe	& 2011-08-24 		& 2004-03-19 \\
    5309	& 71.68		& (14:03:12.90, +54:20:55.60)	& 2011fe	& 2011-08-24 		& 2004-03-14 \\
    4733	& 25.13		& (14:03:12.90, +54:20:55.60)	& 2011fe	& 2011-08-24 		& 2004-05-07 \\
    5322	& 65.53		& (14:03:12.90, +54:20:55.60)	& 2011fe	& 2011-08-24 		& 2004-05-03 \\
    5323	& 43.16		& (14:03:12.90, +54:20:55.60)	& 2011fe	& 2011-08-24 		& 2004-05-09 \\
    4734	& 35.93		& (14:03:12.90, +54:20:55.60)	& 2011fe	& 2011-08-24 		& 2004-07-11 \\
    5337	& 10.07		& (14:03:12.90, +54:20:55.60)	& 2011fe	& 2011-08-24 		& 2004-07-05 \\
    5338	& 28.93		& (14:03:12.90, +54:20:55.60)	& 2011fe	& 2011-08-24 		& 2004-07-06 \\
    5339	& 14.51		& (14:03:12.90, +54:20:55.60)	& 2011fe	& 2011-08-24 		& 2004-07-07 \\
    5340	& 55.12		& (14:03:12.90, +54:20:55.60)	& 2011fe	& 2011-08-24 		& 2004-07-08 \\
    4735	& 29.15		& (14:03:12.90, +54:20:55.60)	& 2011fe	& 2011-08-24 		& 2004-09-12 \\
    6114	& 67.05		& (14:03:12.90, +54:20:55.60)	& 2011fe	& 2011-08-24 		& 2004-09-05 \\
    6115	& 36.2		& (14:03:12.90, +54:20:55.60)	& 2011fe	& 2011-08-24 		& 2004-09-08 \\
    6118	& 11.61		& (14:03:12.90, +54:20:55.60)	& 2011fe	& 2011-08-24 		& 2004-09-11 \\
    4736	& 78.34		& (14:03:12.90, +54:20:55.60)	& 2011fe	& 2011-08-24 		& 2004-11-01 \\
    6152	& 44.66		& (14:03:12.90, +54:20:55.60)	& 2011fe	& 2011-08-24 		& 2004-11-07 \\
    4737	& 22.13		& (14:03:48.20, +54:21:41.00)	& 2011fe	& 2011-08-24 		& 2005-01-01 \\
    6169	& 29.75		& (14:03:48.20, +54:21:41.00)	& 2011fe	& 2011-08-24 		& 2004-12-30 \\
    6170	& 48.56		& (14:03:48.20, +54:21:41.00)	& 2011fe	& 2011-08-24 		& 2004-12-22 \\
    6175	& 41.18		& (14:03:48.20, +54:21:41.00)	& 2011fe	& 2011-08-24 		& 2004-12-24 \\
  \hline
\end{tabular} \label{Table:Observations}
\end{minipage}
\end{table*}

\begin{table*}
 \begin{minipage}{1.0\textwidth}
  \caption{Upper limit bolometric luminosities of nearby ($<$ 25 Mpc) SNe Ia with pre-explosion images.}
  \centering
  \begin{tabular}{@{} c c c c c c c @{}}
  \hline
   Supernova	& position	 	& pre-explosion		& exposure	& counts	& value of	& unabsorbed 3-$\sigma$ \\
		&[RA, DEC]		& \textit{Chandra}	& time		& in		& exposure	& upper limit \\
		&			& observations		& [ks]		& source	& map at	& bolometric \\
		&			&			&		& region	& position	& luminosity\footnote{for 30 eV, 50 eV, 100 eV, \& 150 eV, respectively.}\\
		&			&			&		&		& [s$\cdot$cm$^{2}$]	& [erg/s] \\
 \hline
 \hline
 2002cv	& (10:18:03.68, +21:50:06.0) 	& 2760			& 20.07	& 1		& 4.99481$\cdot10^{6}$	& 3.9$\cdot10^{47}$ \\
	& 				& 			& 	& 1		& 8.85503$\cdot10^{6}$	& 1.7$\cdot10^{44}$ \\
	& 				& 			& 	& 1		& 1.02031$\cdot10^{7}$	& 4.7$\cdot10^{41}$ \\
	& 				& 			& 	& 1		& 1.04034$\cdot10^{7}$	& 1.2$\cdot10^{41}$ \\

\hline
 2003cg & (10:14:15.97, +03:28:02.5)	& 1614 			& 2.15	& 1		& 261570	& 1.2$\cdot10^{43}$ \\
	&				&			&	& 1		& 376688	& 1.8$\cdot10^{41}$ \\
	& 				& 			&	& 1		& 605926	& 7.8$\cdot10^{39}$ \\
	& 				& 			&	& 1		& 740782	& 3.5$\cdot10^{39}$ \\

\hline
 2004W	& (12:43:36.52, +11:31:50.8)	& 785			& 37.35	& 1		& 6.36082$\cdot10^{6}$	& 8.5$\cdot10^{40}$ \\
	&				&			& 	& 1		& 8.75898$\cdot10^{6}$	& 1.8$\cdot10^{39}$ \\
	&				&			& 	& 1		& 1.29863$\cdot10^{7}$	& 1.2$\cdot10^{38}$ \\
	&				&			& 	& 1		& 1.54893$\cdot10^{7}$	& 6.0$\cdot10^{37}$ \\

\hline
 2006X	& (12:22:53.99, +15:48:33.1)	& 400			& 2.53	& 0		& 474548	& 3.2$\cdot10^{42}$ \\
	&				&			&	& 0		& 619526	& 6.6$\cdot10^{40}$ \\
	&				&			&	& 0		& 855029	& 4.3$\cdot10^{39}$ \\
	&				&			&	& 0		& 990341	& 2.2$\cdot10^{39}$ \\

\hline
 2006dd	& (03:22:41.62, -37:12:13.0)	& 2022			& 30.23	& 6		& 2.50339$\cdot10^{6}$	& 1.8$\cdot10^{40}$ \\
	&				&			&	& 6		& 3.22214$\cdot10^{6}$	& 1.1$\cdot10^{39}$ \\
	&				&			&	& 6		& 5.55312$\cdot10^{6}$	& 1.4$\cdot10^{38}$ \\
	&				&			&	& 6		& 7.48767$\cdot10^{6}$	& 9.1$\cdot10^{37}$ \\

\hline
 2006mr	& (03:22:42.84, -37:12:28.5)	& 2022			& 30.23	& 44		& 2.43645$\cdot10^{6}$	& 6.9$\cdot10^{40}$ \\
	&				&			&	& 44		& 3.14788$\cdot10^{6}$	& 3.5$\cdot10^{39}$ \\
	&				&			&	& 44		& 5.46562$\cdot10^{6}$	& 5.1$\cdot10^{38}$ \\
	&				&			&	& 44		& 7.39541$\cdot10^{6}$	& 3.4$\cdot10^{38}$ \\

\hline
 2007gi	& (12:01:23.42, +61:53:33.8)	& 6783			& 15.13	& 0		& 444218		& 2.1$\cdot10^{41}$ \\
	&				&			&	& 0		& 743225		& 6.8$\cdot10^{39}$ \\
	&				&			&	& 0		& 1.76997$\cdot10^{6}$	& 5.5$\cdot10^{38}$ \\
	&				&			&	& 0		& 2.63497$\cdot10^{6}$	& 3.0$\cdot10^{38}$ \\

\hline
 2007sr	& (12:01:52.80, -18:58:21.7)	& 315, 3040, 3041, 3042,& 425.18\footnote{total exposure time of the combined image.} & 4		& 1.03024$\cdot10^{7}$	& 9.3$\cdot10^{39}$ \\
	&				& 3044, 3044, 3718	&	& 4		& 1.57593$\cdot10^{7}$	& 3.7$\cdot10^{38}$ \\
	&				&			&	& 4		& 3.61044$\cdot10^{7}$	& 3.7$\cdot10^{37}$ \\
	&				&			&	& 4		& 5.52568$\cdot10^{7}$	& 2.1$\cdot10^{37}$ \\

\hline
 2008fp	& (07:16:32.60, -29:19:31.7)	& 4866			& 30.16	& 1		& 1.79972$\cdot10^{6}$	& 1.0$\cdot10^{42}$ \\  
	&				&			&	& 1		& 3.07844$\cdot10^{6}$	& 1.7$\cdot10^{40}$ \\
	&				&			&	& 1		& 6.00491$\cdot10^{6}$	& 8.2$\cdot10^{38}$ \\
	&				&			&	& 1		& 7.93188$\cdot10^{6}$	& 3.7$\cdot10^{38}$\\

\hline
 2011fe	& (14:03:05.81, +54:16:25.4)	& 4731, 5296, 5297, 5300,& 898.17\footnote{total exposure time of the combined image.} & 4		& 7.63589$\cdot10^{6}$ / 7.37164$\cdot10^{6}$		& 6.5$\cdot10^{38}$ / 4.4$\cdot10^{38}$ \\
	&				& 4732, 5309, 4733, 5322,&	& 4		& 1.33861$\cdot10^{7}$ / 1.23642$\cdot10^{7}$	& 2.4$\cdot10^{37}$ / 1.8$\cdot10^{37}$ \\
	&				& 5323, 4737, 5338, 5339,&	& 4		& 3.87832$\cdot10^{7}$ / 3.5352$\cdot10^{7}$	& 2.0$\cdot10^{36}$ / 1.7$\cdot10^{36}$ \\
	&				& 5340, 4735, 6114, 6115,&	& 4		& 6.49251$\cdot10^{7}$ / 6.00846$\cdot10^{7}$	& 1.0$\cdot10^{36}$ / 9.3$\cdot10^{35}$ \\
	&				& 6118, 4736, 6152, 4737,&	&		& 		& \\
	&				& 6169, 6170, 6175	&	&		& 		& \\

\hline
\end{tabular} \label{Table:Upper.Limits}
\end{minipage}
\end{table*}

\newpage
\clearpage

\newpage
\clearpage

\label{lastpage}

\end{document}